# Mn-induced magnetic symmetry breaking and its correlation with the metal-insulator transition in bilayered $Sr_3(Ru_{1-x}Mn_x)_2O_7$

Qiang Zhang[1,2], Feng Ye[2], Wei Tian[2], Huibo Cao[2], Songxue Chi[2], Biao Hu[1], Zhenyu Diao[1], David A. Tennant,[2] Rongying Jin[1], Jiandi Zhang[1], and Ward Plummer[1]

[1]*Department of Physics and Astronomy, Louisiana State University, Baton Rouge, LA 70803*
[2]*Oak Ridge National Laboratory, Oak Ridge, Tennessee 37831, USA*

## Abstract

Bilayered $Sr_3Ru_2O_7$ is an unusual metamagnetic metal with inherently antiferromagnetic (AFM) and ferromagnetic (FM) fluctuations. Partial substitution of Ru by Mn results in the establishment of metal-insulator transition (MIT) at $T_{MIT}$ and AFM ordering at $T_M$ in $Sr_3(Ru_{1-x}Mn_x)_2O_7$. Using elastic neutron scattering we determined the effect of Mn doping on the magnetic structure and in-plane magnetic correlation lengths in $Sr_3(Ru_{1-x}Mn_x)_2O_7$ ($x$ = 0.06 and 0.12). With increasing Mn doping ($x$) from 0.06 to 0.12, an evolution from an in-plane short-range to long-range antiferromagnetic (AFM) ground state occurs. For both compounds, the magnetic ordering has the double-stripe configuration, with the onset of correlation coinciding with the sharp rise of the electrical resistivity and the specific heat. Since it does not induce measurable lattice distortion, the double-stripe magnetic order with anisotropic spin texture breaks the symmetry from $C_{4v}$ crystal lattice to $C_{2v}$ magnetic sublattice. These observations shed new light on an age-old question of Slater versus Mott-type MIT.

Bilayered Ruthenate $Sr_3Ru_2O_7$ and its derivatives have attracted considerable attention due to their intriguing physical properties resulting from the interplay between charge, spin, orbital and lattice degrees of freedom. The undoped $Sr_3Ru_2O_7$ (Fig. 1(a)) is a paramagnetic metal with the electrical conduction in the $RuO_2$ layers [1] and no long-range magnetic order, although magnetic susceptibility exhibits a peak at ~18 K [2]. Instead, $Sr_3Ru_2O_7$ shows two-dimensional (2D) ferromagnetic (FM) fluctuations above 18 K but incommensurate antiferromagnetic (AFM) fluctuations below 18 K [3]. The competing FM and AFM interactions lead to the metamagnetic transition at low temperatures accompanied with quantum critical behavior at the transition field $H_c$ ($H // c$, $c$ defined in Fig. 1(a)) ~ 8 Tesla [4-8]. In the vicinity of $H_c$, there is a very strong increase of the electrical resistivity with large anisotropy [9], lattice distortion [10], and spin-density-wave (SDW) phase with propagation vector $q$ = (0.233 0 0) along the $Ru$-$O$-$Ru$ bond direction [11]. These observations indicate that the four-fold rotation symmetry ($C_4$) is reduced to two-fold ($C_2$), giving rise to so-called electronic nematic fluid below $T_{nem} \approx 1$ K [9-15]. In addition to magnetic field, the electrical resistivity of the anomalous phase strongly responds to other external forces such as in-plane uniaxial stain [16]. An intriguing question is: would an internal pressure induced by ion substitution also break the symmetry?

With the substitution of $4d$ Ru using more localized $3d$ Mn, the resultant $Sr_3(Ru_{1-x}Mn_x)_2O_7$ reveals drastically different electronic and magnetic properties from the undoped compound [17-20]. With $x > 0.06$, $Sr_3(Ru_{1-x}Mn_x)_2O_7$ exhibits the metal-insulator transition at $T_{MIT}$ (the electrical resistivity slope changes sign) and antiferromagnetic (AFM) magnetic ordering at $T_M$ [17-18, 20], where $T_M$ is determined from the peak in the magnetic susceptibility. Using x-ray absorption spectroscopy and resonant elastic x-ray scattering,

Hossain et al. [20] investigated the nature of MIT in $Sr_3(Ru_{1-x}Mn_x)_2O_7$ and concluded that the MIT is Mott (electronic correlations) rather than Slater (AFM ordering) type. While $T_{MIT}$ increases with increasing $x$, $T_M$ reaches its maximum at $x \sim 0.16$, where the $RuO_6$ rotation angle approaches zero [17]. The magnetic structure of $Sr_3(Ru_{1-x}Mn_x)_2O_7$ ($x = 0.16$) with the highest $T_M$ of 81 K has been determined to be so-called "$E$-type" AFM ordering [21], as shown in Fig. 1(c). The moments form long-range order in the $ab$ plane with short magnetic correlation length along the $c$ axis (5 - 6 Å), reflecting a quasi-2D character. The in-plane spin arrangements are "up-up-down-down" along the sides and antiparallel/parallel along the diagonal direction of (Ru/Mn)-(Ru/Mn) square lattice. Such a magnetic configuration is surprising in the context of a square lattice structure. Emergent phenomena induced by Mn doping raise a few interesting questions: 1) How does the magnetic correlation length evolve with Mn doping, starting with fluctuations in the parent compound? 2) How can the $E$-type magnetic order form in a square lattice? 3) What is the relationship between the MIT and the magnetic transition? Using elastic neutron scattering, we show that substituting Mn for Ru in $Sr_3(Ru_{1-x}Mn_x)_2O_7$ stabilizes the long-range AFM order with the double - stripe magnetic structure in the $ab$ plane ($C_{2v}$), while the crystal lattice remains the $C_{4v}$ symmetry. For the three Mn concentrations studied, the onset of magnetic correlation coincides with the sharp rise of the electrical resistivity, indicating the Slater-like MIT transition [22], that is, a MIT due to the formation of magnetic ordering. These features and the anisotropy in the magnetic correlation lengths set $Sr_3(Ru_{1-x}Mn_x)_2O_7$ as a unique magnetic system, indicative of a magnetic crystal.

The $Sr_3(Ru_{1-x}Mn_x)_2O_7$ ($x = 0.06$ and 0.12) (denoted as SRMO6 and SRMO12 hereafter) crystals were grown using the floating-zone technique, and were well characterized [17].

Fig. 1(b) shows the temperature dependence of the in-plane magnetic susceptibility ($\chi_{ab}$) for SRMO6 and SRMO12. For comparison, $\chi_{ab}$ for $x = 0.16$ is also presented. Aligned SRMO6 and SRMO12 single crystals were used for the elastic neutron diffraction at the fixed-incident-energy (wavelength $\lambda = 2.365$ Å) triple-axis spectrometer HB1A and four-circle single crystal diffractometer with a wavelength of 1.542 Å (which includes ~ 1.4% $\lambda/2$ contamination) at the High Flux Isotope Reactor (HFIR) Oak Ridge National Laboratory, USA.

Similar to undoped $Sr_3Ru_2O_7$ [23,24], the structure of $Sr_3(Ru_{1-x}Mn_x)_2O_7$ ($x = 0.06, 0.12$ and 0.16) at room temperature (RT) is orthorhombic with space group *Bbcb*. The orthorhombicity results from the octahedral rotation, thus yielding the same in-plane lattice constants (i.e., $a_o = b_o$). Figure 1(a) shows the tetragonal unit cell for $Sr_3(Ru_{1-x}Mn_x)_2O_7$ where (Ru/Mn) ions form a square lattice with (Ru/Mn)-O-(Ru/Mn) bond directions pointing along the tetragonal crystalline axes ($a_T$ and $b_T$). Due to the octahedral rotation, the real unit cell is orthorhombic with $a_o = b_o = \sqrt{2}a_T$ as marked by the green square in Fig. 1(d). The *E*-type magnetic unit cell is a (2×1) super cell, represented by the orange rectangle in Fig. 1(d). Within the neutron resolution in HB3A, no difference between lattice parameters $a_o$ and $b_o$ is observed for SRMO6 and SRMO12 between RT and 5 K. The lattice constants are $a_o = b_o = 5.4885(4)$ Å, $c = 20.5277(6)$ Å for SRMO12 and $a_o = b_o = 5.4711(3)$ Å, $c = 20.7146(5)$ for SRMO6 at 5 K, with 0.2% change in volume.

Figures 2(a) and 2(b) show, at 2 K, the longitudinal *H* scan along the $a_o$ axis for SRMO6, SRMO12, and SRMO16 ($x = 0.16$), and the transverse *K* scan along the $b_o$ axis for SRMO6 and SRMO12 through the AFM propagation vector $\mathbf{Q_{AFM}} = (0.5\ 0\ 0)$, respectively. The longitudinal *H* scans for SRMO12 and SRMO16 overlap each other and are resolution limited, indicating that the $x = 0.12$ compound behaves similarly to $x = 0.16$ with the in-plane long-range AFM order at low temperatures. However, both the longitudinal and transverse scans for SRMO6 at 2 K are broader than that of SRMO12. The broader magnetic peaks seen in SRMO6 cannot be related to the differences in the sample quality/crystallinity, since it has similar mosaic as that of SRMO12. We thus conclude the broader peaks are the reflection of the intrinsic magnetism, which indicates SRMO6 forms only short-range but not long-range AFM order in the *ab* plane at 2 K.

Magnetic structure refinement on $Sr_3(Ru_{1-x}Mn_x)_2O_7$ ($x = 0.06$ and $0.12$) at 5 K reveals a double-stripe AFM order with the following features: (1) The ordered moment on $(Ru/Mn)_1$ and $(Ru/Mn)_2$ sites is the same and points along the *c* axis (see Fig. 1(c)); (2) In the *ab*-plane, spins form two interpenetrating stripe-like AFM ordering with double ferromagnetic stripes along the $b_o$ direction, alternating antiferromagnetically along the $a_o$ direction (see Fig. 1(d)); (3) The spin texture is the same for both layers within the unit cell (see Fig. 1(c)). As $x$ reduces from 0.16 to 0.06, the spin configuration at 5 K remains similar but there is a crossover from the long-range to the short-range order. Furthermore, the ordered moment of Ru/Mn becomes smaller, from 0.70 $\mu_B$ for SRMO16, 0.51$\mu_B$ for SRMO12 to 0.18$\mu_B$ for SRMO6.

The temperature (*T*) dependence of the longitudinal *H* (along the $a_o$ direction) and transverse *K* scans (along the $b_o$ direction) through $\mathbf{Q}_{AFM}$ in SRMO6 and SRMO12 is measured. Figs. 2(c) and 2(d) show these scans at $T = 4$ K and 65 K for SRMO12. While both *H* and *K* scans are resolution limited at 4 K, the peak is clearly broadened at 65 K, consistent with the resonant x-ray scattering results [20], which provides direct evidence of the short-range order above $\approx 60$ K. Similar trend is also observed in SRMO6 as presented in Figs. 2(e) and 2(f), which show the H and K scans at T = 2 K and 22 K. Note the broadening is also seen above ≈ 18 K, indicating that the short-range order in SRMO6 extends above $T_M$. The magnetic peaks eventually disappear at $T_{AF}^{onset}$ ~ 24 K for SRMO6 and ~ 70 K for SRMO12 when T approaches $T_{MIT}$, which is ~ 24 K for SRMO6 and ~ 90 K for SRMO12 [17]. Figs. 3(a) and 3(b) show the temperature dependence of the full-width-at-half-maximum (FWHM) for both longitudinal H scans along the $a_o$ axis and transverse K scans along the $b_o$ axis obtained from the fits to the data using Gaussian function for SRMO6 and SRMO12, respectively. The short-range order with relatively high FWHM starts to be detected at $T_{AF}^{onset}$ > $T_M$. As T decreases, FWHM decreases gradually and appears to be saturated at around 18 K and 60 K for SRMO6 and SRMO12, respectively.

In order to determine the intrinsic correlation lengths along the $a_o$ and $b_o$ directions we fit the experimental line-shape of the longitudinal and transverse scans to a Lorentz function $L(q) = \frac{c}{1+(q \times \xi)^2}$, convoluted with the instrumental resolution Gaussian function $G(q) = \exp(-q^2/(2\left(\frac{\sigma}{2\sqrt{ln4}}\right)^2))$ [26]. Here, $\xi$ ($\xi^1$ is half width at half maximum of Lorentzian function) is defined as the magnetic correlation length [27], and σ is the FWHM of instrumental resolution Gaussian function with $\sigma_{ao}$ ≈ 0.0353 Å$^{-1}$ and $\sigma_{bo}$ ≈ 0.0127 Å$^{-1}$ (see two horizontal dashed lines in Figs. 3(a) and (b)) determined by *Reslib* program [25]. The deconvolution yields the magnetic correlation lengths $\xi_{ao}$ and $\xi_{bo}$ plotted in Figures 3(c) and (d). The onset for short-range magnetic order is marked using $T_{AF}^{onset}$ for both SRMO6 and SRMO12. As the temperature is lowered, both $\xi_{ao}$ and $\xi_{bo}$ increase with anisotropic behavior ($\xi_{ao}$ > $\xi_{bo}$). Below $T_M$, the magnetic correlation saturates for SRMO06 with $\xi_{a\text{o}} \sim 1.5\xi_{b\text{o}}$, confirming the short-range order. The data for SRMO12 also shows anisotropic behavior below $T_{AF}^{onset}$, with $\xi_{a\text{o}} > \xi_{b\text{o}}$. True long-range magnetic order is evident in $\xi_{b\text{o}}$ below ≈ 60 K ($<T_\text{M}$). Unfortunately, we cannot determine $\xi_{a\text{o}}$ below ≈ 60 K because the error bar becomes too large. It should be noted that the anisotropic behavior observed in both compositions is counter-intuitive, since spin structure breaks the crystal symmetry (Fig. 1(c)). One would expect that the lattice would try to destroy the magnetic crystallization.

It would be of importance to address the magnetic interactions and the origin of the anisotropic magnetic correlation length in the double-stripe order in $Sr_3(Ru_{1-x}Mn_x)_2O_7$. To stabilize each of the two stripes, AFM next-nearest-neighbour (NNN) coupling should be larger than half of the AFM nearest-neighbour (NN) coupling within each stripe ordered sublattice, as reported in Fe-based pnictides with the single-stripe order [28] and FeTe with the double stripe order [29]. Since the NN and NNN intracoupling within each stripe ordered sublattice are AFM couplings, the NN bonds along the spin-parallel ($b_\text{o}$) direction are frustrated as compared to the NN bonds along the spin-antiparallel ($a_\text{o}$) direction, resulting in $\xi_{a\text{o}} > \xi_{b\text{o}}$. Interestingly, unlike the single-stripe order case with two degenerate magnetic stripe states ($\pi$, 0) and (0, $\pi$) [27, 28], the double-stripe AFM order in $Sr_3(Ru_{1-x}Mn_x)_2O_7$ has four degenerate magnetic stripe states with spin frustration, as depicted in Figs. 3(e-h).

Figure 4(a) shows the temperature dependence of the integrated intensities for the magnetic peak (0.5 0 0) for three different Mn concentrations ($x$ = 0.06, 0.12, and 0.16). Note that the onset of magnetic order occurs prior to $T_\text{M}$, i.e., $T_{AF}^{onset} > T_\text{M}$. Interestingly, for every Mn concentration studied, there is a dramatic increase in the resistivity ($T_\rho^{rise}$), much more pronounced than that at $T_{MIT}$. This is clearly demonstrated in the inset of Fig. 4(a): at the onset of specific heat anomaly (the same as $T_{AF}^{onset}$), there is a sharp rise in the resistivity, i.e., $T_{AF}^{onset} = T_{\rho}^{rise}$. Fig. 4(b) summarizes the temperature and Mn-content ($x$) dependence of the magnetism. Regions I-II denote the paramagnetic metallic (PM-M) and paramagnetic insulating (defined by dρ/dT < 0) state, respectively. Region III is the short-range double-stripe ordered insulating state (SR-I), and Region IV is the long-range double-stripe ordered insulating state (LR-I). The magnetic structure in Region III is schematically shown in the inset of Fig. 4(b). We emphasize three important results in our phase diagram. First, there is anisotropy in the magnetic correlation length, i.e., $\xi_{ao} > \xi_{bo}$ in Regions III, which is related to the anisotropic spin texture. Second, the short-range double-stripe AFM order (Region III) becomes detectable below $T_{AF}^{onset}$ (>$T_M$), and a huge resistivity response always accompanies the onset of magnetic correlation at $T_{AF}^{onset}$, indicating that $Sr_3(Ru_{1-x}Mn_x)_2O_7$ is a Slater-like insulator. The long-range magnetic order develops upon cooling as reflected by enhanced magnetic correlation lengths (Region IV). Third, given that single-crystal x-ray diffraction shows no structure change across $T_{MIT}$ and $T_M$, $Sr_3(Ru_{1-x}Mn_x)_2O_7$ ($x$ = 0.06, 0.12, 0.16) should have the same structural symmetry as the undoped sample [3], with the fourfold symmetry ($C_{4v}$) in the $ab$ plane [17]. The double-stripe magnetic structure below $T_{AF}^{onset}$ has twofold symmetry ($C_{2v}$) due to the anisotropic spin texture.

To understand our experimental results, we compare the Mn - induced magnetism with that induced by magnetic field. Mn doping induces the double-stripe AFM order with the propagation vector (0.5 0 0) along the diagonal direction of (Ru/Mn)-(Ru/Mn) square lattice. In contrast, magnetic field induces SDW with the propagation vector (0.233 0.233

0) and resistivity anisotropy due to the nematic order along the side of the Ru-Ru square lattice of $Sr_3Ru_2O_7$ [11]. In spite of these differences, both Mn doping and magnetic field result in the reduced symmetry in physical properties. In the case of $Sr_3(Ru_{1-x}Mn_x)_2O_7$, the symmetry breaking occurs below $T_{AF}^{onset}$. If the nematic order emerges below $T_{nem}$ (=$T_{AF}^{onset}$) in the Region III, it could act as coupling to lock the nearest-neighbor spins within individual stripe ordered sublattice in a ferromagnetic- or antiferromagnetic-like configuration [28,30], and therefore the frustration resulting from four degenerate magnetic stripe states present at higher temperatures can be lifted, leading to symmetry breaking.

In summary, we have investigated the Mn-induced magnetism through elastic neutron scattering in bilayered $Sr_3(Ru_{1-x}Mn_x)_2O_7$. With increasing Mn concentration, the AFM interaction gradually increases, forming initially short-range double-stripe AFM order then long-range order in the *ab* plane. For both $x = 0.06$ and 0.12, the magnetic correlation becomes detectable at $T_{AF}^{onset} > T_M$. The double-stripe AFM configuration presents $C_{2v}$ symmetry, which is lower than the crystal symmetry ($C_{4v}$). Anisotropic magnetic correlation length with $\xi_{ao} > \xi_{bo}$ is found in such spin texture. The symmetry breaking may reflect complex yet interesting underlying physics such as magnetic crystal formation.

The measurement of magnetic correlation enables us to understand the interplay between magnetism, symmetry breaking, and MIT in $Sr_3(Ru_{1-x}Mn_x)_2O_7$. Due to inherent magnetic instability in $Sr_3Ru_2O_7$, the introduction of magnetic Mn with narrower 3*d* band (compared to Ru) stabilizes the AFM interaction and result in both magnetic and MIT transitions. With the low Mn concentration, we observe that the magnetic correlation becomes measureable at MIT indicating Slater-like spin-charge coupling. While the MIT occurs at higher temperature than the magnetic correlation temperature as *x* increases, there is always a sharp resistivity rise, and a specific heat anomaly at the onset of magnetic order, indicating that the transition is driven by spin, instead of charge correlation or structural change. The separation between the MIT and the true magnetic transition suggests that the electron-electron interaction continuously increases with increasing $x$, which may eventually drive the system to Mott-type insulator as proposed by *Ref.* [20].

**Acknowledgments**

We would like to thank Mohammad Saghayezhian with his assistance using the Voight function in *OriginLab* software to fit the data and Dalgis Mesa for sharing with us unpublished data from her thesis. Primary support for this study came from the U.S. Department of Energy under EPSCoR Grant No. DE-SC0012432 with additional support from the Louisiana Board of Regents. Use of the high flux isotope reactor at the Oak Ridge National Laboratory, was supported by the US Department of Energy, office of Basic Energy Sciences, Scientific User Facilities Division.

Figure 1: (a) A 3D view of the tetragonal unit cell for $Sr_3(Ru_{1-x}Mn_x)_2O_7$; (b) Temperature dependence of the in-plane magnetic susceptibility $\chi_{ab}$ for $Sr_3(Ru_{1-x}Mn_x)_2O_7$ ($x = 0.06$, $0.12$, and $0.16$) measured by applying 1000 Oe magnetic field parallel to the $a_o b_o$ plane; (c) The schematic magnetic structure of $Sr_3(Ru_{1-x}Mn_x)_2O_7$ ($x = 0.06$ and $0.12$) obtained from the refinement of our neutron diffraction data at 4 K, where $a_o$ and $b_o$ are the axes of the orthorhombic unit cell; (d) The in-plane double-stripe AFM spin structure of $Sr_3(Ru_{1-x}Mn_x)_2O_7$. The solid green square shows the orthorhombic unit cell with Ru/Mn atoms at the center of each edge. The black dotted square indicates the tetragonal unit cell with Ru/Mn atoms at the corners. The orange rectangle shows the (2×1) magnetic unit cell.

Figure 2: (a) Longitudinal $H$ scans along the $a_o$ axis for $x = 0.06$, $0.12$ and $0.16$ at 2 K; (b) Transverse $K$ scans along the $b_o$ axis for $x = 0.06$ and $0.12$ at 2 K. The solid curves are the fits to the data using the Gaussian function to compare the FWHM among different compounds at 2 K. The longitudinal $H$ scans and transverse $K$ scans are illustrated by green arrows in the insets of (a) and (b), respectively; (c) Longitudinal $H$ scans along the $a_o$ axis, and (d) transverse $K$ scans along the $b_o$ axis through $\mathbf{Q}_{AFM} = (0.5\ 0\ 0)$ at 4 K (below $T_M$) and 65 K ($\sim T_M$) for $x = 0.12$; (e) Longitudinal $H$ scans along the $a_o$ axis, and (f) transverse $K$ scans along the $b_o$ axis through $\mathbf{Q}_{AFM} = (0.5\ 0\ 0)$ at 2 K (below $T_M$) and 22 K ($\sim T_M$) for $x = 0.06$. The solid lines in (c)-(f) are the fits to the data using our model to obtain the intrinsic magnetic correlation length as described in the main text. The instrumental resolution is shown as the bars.

Figure 3: Temperature dependence of the FWHM along the $a_o$ and $b_o$ axes for $x = 0.06$ (a) and $x = 0.12$ (b) obtained from the fits to the data using the Gaussian function. The horizontal dashed blue and red lines indicate the instrumental resolution $\sigma_{ao}$ and $\sigma_{bo}$, respectively. (c) and (d) show the temperature dependence of the magnetic correlation length $\xi_{ao}$ and $\xi_{bo}$ for $x = 0.06$ and $x = 0.12$, respectively; (e-h) Four degenerate magnetic states of the double-stripe AFM order in $Sr_3(Ru_{1-x}Mn_x)_2O_7$. The two dashed lines indicate the directions of the double ferromagnetic stripes.

Figure 4: (a) Temperature dependence of the integrated intensities of (0.5 0 0) magnetic peak obtained from the fits to their longitudinal scans along the $a_o$ axis using the Gaussian function for $x = 0.06$, $0.12$ and $0.16$. The inset shows the temperature dependence of specific heat plotted as $C_p/T$ and resistivity for $x = 0.16$, demonstrating the sharp resistivity rise at the onset of specific heat anomaly; (b) Phase diagram of $Sr_3(Ru_{1-x}Mn_x)_2O_7$ constructed by combining elastic neutron scattering results with previously published data [17]: Region I is the paramagnetic metallic state, Region II is the paramagnetic insulating state defined by $d\rho/dT < 0$, Region III is the short-range AFM insulating state, and Region IV is the long-range AFM insulating state. The inset is the short-range double-stripe AFM structure in Region III.

Figure 1.

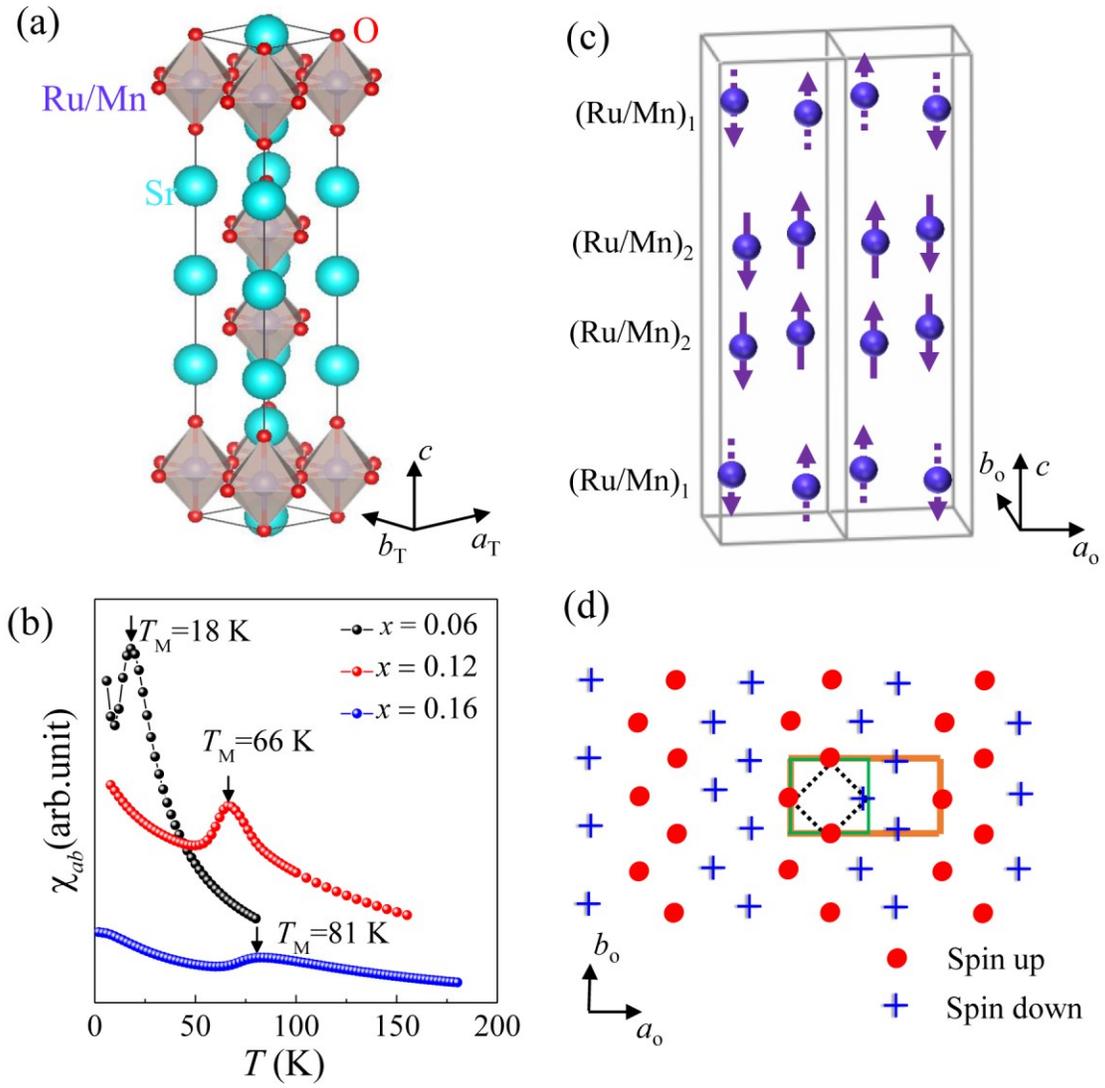

Figure 2.

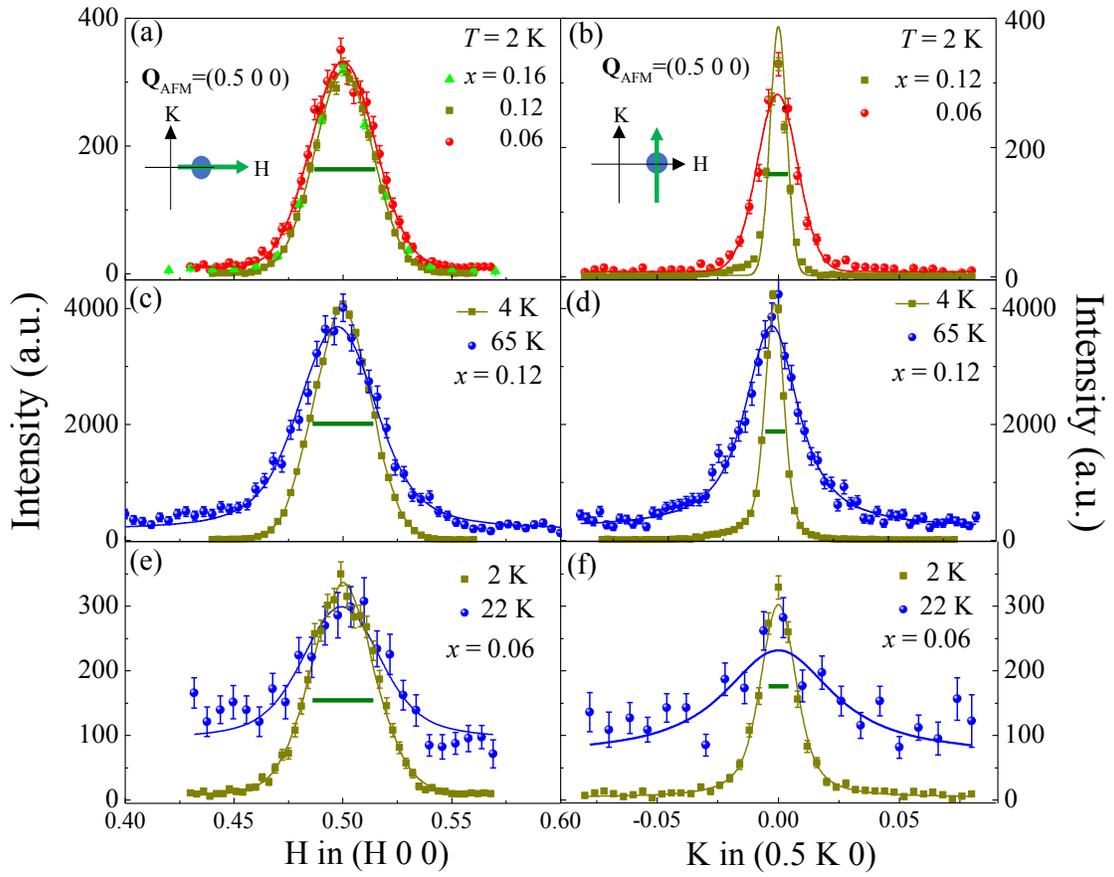

Figure 3.

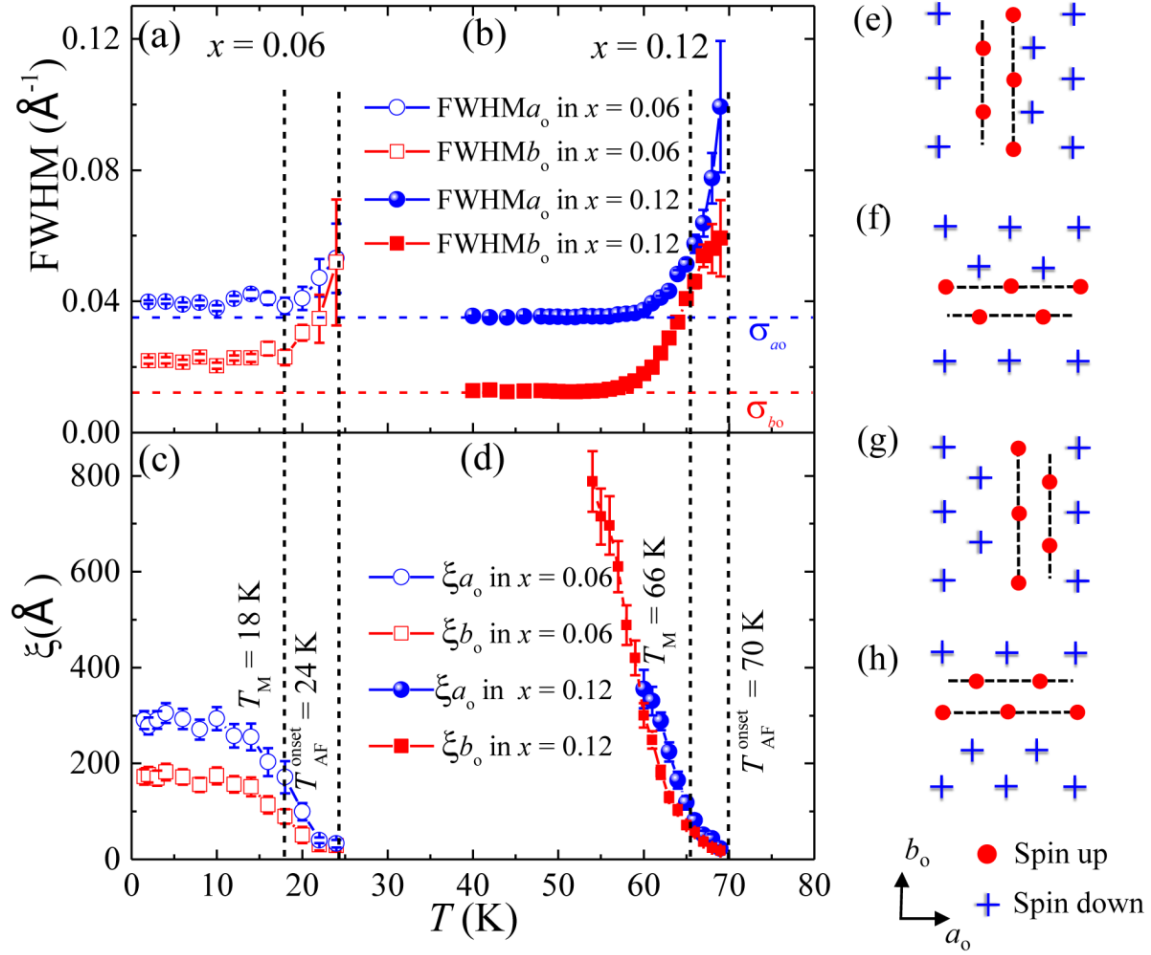

Figure 4.

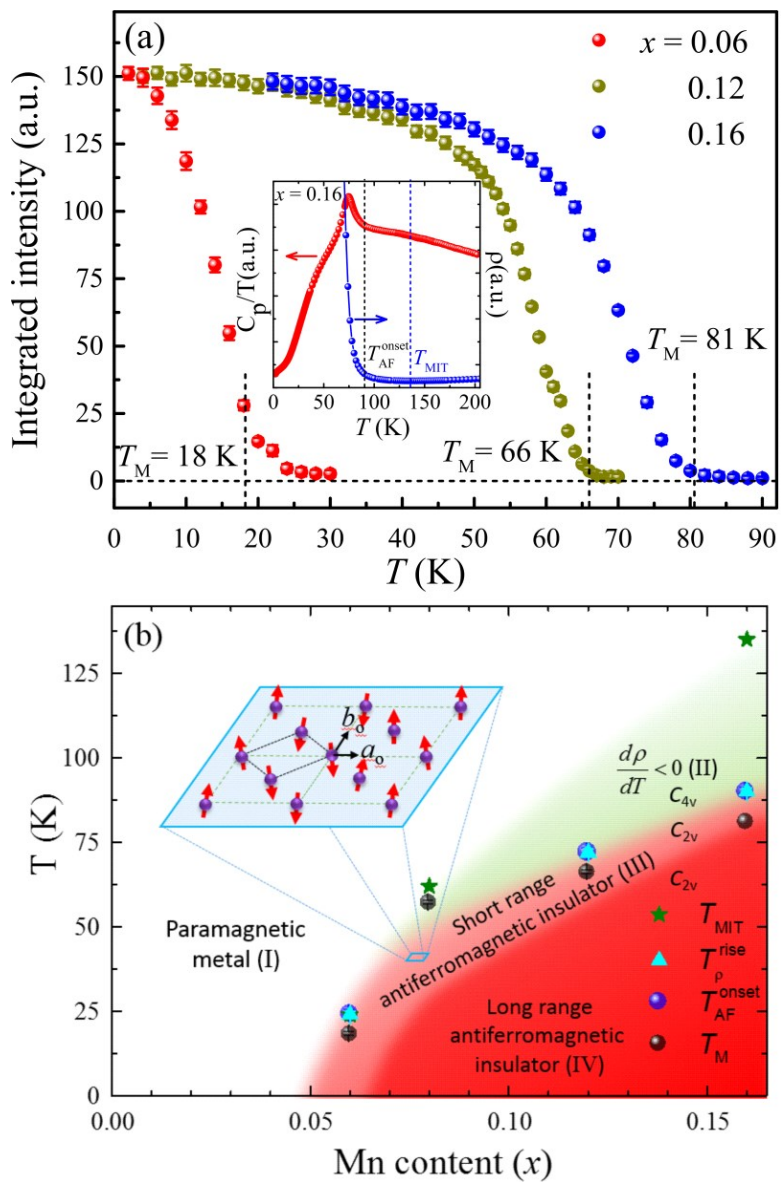